\documentclass[aip,apl,reprint,twocolumn]{revtex4-1}	   % 2spaltig

\usepackage{natbib}
\usepackage{graphicx}% Include figure files

\begin{document}

\title{Catalytic Growth of N-doped MgO on Mo(001)} %Title of paper

\author{Martin Grob}\email[corresponding author: ]{grob@physik.rwth-aachen.de}
\author{Marco Pratzer}
\affiliation{II. Institute of Physics B and JARA-FIT, RWTH Aachen University, D-52074 Aachen, Germany}
\author{Marjana Le\v{z}ai\'{c}}
\affiliation{Peter-Gr\"unberg Institute , Forschungszentrum J\"ulich and JARA, D-52425 J\"ulich, Germany}

\author{Markus Morgenstern}
\affiliation{II. Institute of Physics B and JARA-FIT, RWTH Aachen University, D-52074 Aachen, Germany}

\date{\today}

\begin{abstract}
A simple pathway to grow thin films of N-doped MgO (MgO:N), which has been found experimentally to be a ferromagnetic d$^0$ insulator, is presented. It relies on the catalytic properties of a Mo(001) substrate using growth of Mg in a mixed atmosphere of O$_2$ and N$_2$. Scanning tunneling spectroscopy
reveals that the films are insulating and exhibit an N-induced state slightly below the conduction band minimum.
\end{abstract}

\maketitle %\maketitle must follow title, authors, abstract and \pacs


%Introduction
Recently, it has been found that MgO:N-films grown by molecular beam epitaxy (MBE)
exhibit ferromagnetism after being
annealed at 1020 K.\cite{Yang} The optimized  N-concentration was 2.2 \% exhibiting coercive fields as large as 60 mT at $T=10$ K and magnetic moments per N-atom
of 0.3 $\mu_{\rm B}$  barely reducing up to room temperature. A Curie temperature
$T_{\rm C}\simeq 550$ K has been extrapolated and indications that N is incorporated substitutionally on the O-site
have been deduced from core level spectroscopy. Moreover, independent studies of N implantation (80 keV) into MgO led to a hysteresis
with a coercive field of 30 mT at 300 K.\cite{Ming}
This raises hope that reliable d$^0$ ferromagnetism avoiding d-metals can be realized in MgO:N at $T=300$ K.
Such magnetism without d-orbitals has previously been found in thin films of undoped oxides \cite{HfO2} including MgO \cite{MgO} or defective carbon systems \cite{Esqui}, however, with limited control since relying on defects. ZnO with sp-type dopants as C, N, B, Li, Na, Mg, Al, and Ga shows ferromagnetic signals, too,\cite{Pan,Yu} but, likely, Zn or O vacancies and Zn d-orbitals are involved in the magnetic coupling \cite{Pan,Li}.\\
The d$^0$ ferromagnetism
has first been proposed theoretically relying on the double exchange mechanism in narrow impurity bands.\cite{Kenmochi} But the high $T_{\rm C}$ proposed originally has been challenged by going beyond the mean-field approximation \cite{Mavropoulos}
or by considering correlation effects,\cite{Droghetti, Slipukhina, Gu}. Partly, even the absence of ferromagnetism was found.\cite{Droghetti} This renders the high $T_{\rm C}$ observed experimentally into obvious disagreement to current theory asking for more detailed studies .\\

\begin{figure}[t]
  \includegraphics[width=8cm]{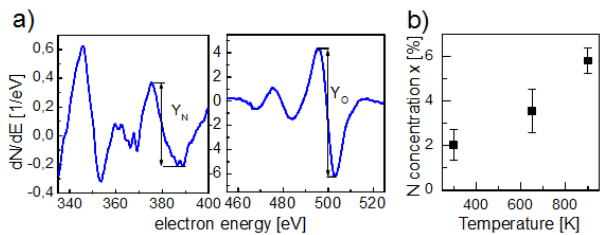}
  \caption{\label{fig_1}(Color online) a) Differential AES spectra of N and O for $x=3.1\%$ with the corresponding peak to peak height $Y_i$ marked. b) Nitrogen concentration $x$ of 7 monolayer thick MgO:N films as a function of deposition temperature $T_{\rm D}$.}
\end{figure}

MgO:N films, in addition, exhibit bipolar resistive switching behaviour \cite{Waser} prior to annealing. Resistance contrasts as large as 4 orders of magnitude, switching currents as low as 100 nA and switching times into both states below 10 ns have been obtained.\cite{Yang} This makes MgO:N also interesting for nonvolatile memories.\\
However, the incorporation of N into MgO is difficult due to the strongly endothermal incorporation of N atoms with respect to N$_2$ (energy cost per N atom: 10 eV) \cite{Pesci}. It requires, e.g., atomic beams of N and  O produced by a high-frequency ion plasma
source\cite{Yang} or N$^+$ implantation \cite{Ming}. Here, we demonstrate a simplified pathway using the catalytic abilities of a Mo(001)
substrate. Thus, we establish a model system of MgO:N for surface science. Mo(001) is chosen since thin MgO films of high quality can be grown epitaxially due to the relatively small lattice mismatch of 6\,\%.\cite{Gallagher, Benedetti, Grob} Moreover, catalytic properties of Mo with respect to N$_2$ are known, as, e.g., for the nitrogenase within bacteria using molybdenum enzymes as catalyst.\cite{Chatt} Catalytic N$_2$
dissociation on surfaces has been induced successfully for the electronically similar W(001), where growth properties of MgO are, however, unknown.\cite{Alducin, Clavenna, Rettner}
We have grown thin films of MgO:N on Mo(001) with thicknesses up to ten monolayer (ML) at optimal doping. Scanning tunneling spectroscopy (STS) revealed that the Fermi level is well within the band gap indicating insulating behaviour of the film. Moreover, an unoccupied state close to the conduction band is found, which is not present in pure MgO films.\\
\begin{figure}
 \includegraphics[width=8cm]{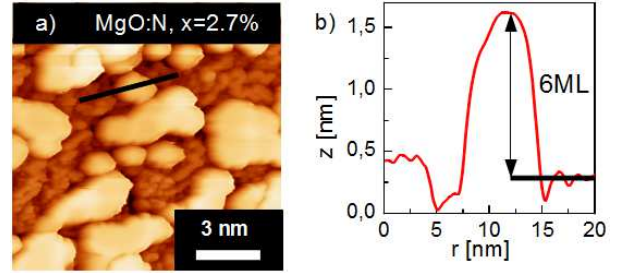}
  \caption{\label{fig_0}(Color online) a) STM image of 7 ML MgO$_{0.973}$N$_{0.027}$ film on Mo(001) with thicker islands (bright areas) on top of a wetting layer of 1-2\,ML; $(50 \times 50)$\,nm$^2$, $U = 3$\,V, $I = 0.5$\,nA. b) Line profile along the black line marked in (a) with height of the island above the wetting layer marked.}
\end{figure}
%
%preparation
The experiments are performed in ultra-high vacuum at a base pressure of $p=5\cdot 10^{-11}$\,mbar.
Firstly, the Mo(100) crystal was cleaned by cyclically annealing within O$_2$ at initial pressure $p_{\rm O_2} = 5\cdot 10^{-7}$\,mbar and 1400\,K, followed by flashing  to 2300\,K. \cite{Bode} After every cycle, $p_{\rm O_2}$ is slightly reduced. The MgO:N films are prepared by molecular beam epitaxy of magnesium at $p_{\rm O_2} = 1\cdot 10^{-7}$\,mbar and N$_2$ pressure $p_{\rm N_2} = 5\cdot 10^{-6}$\,mbar. The deposition temperature $T_{\rm D}$ is 300 K, if not given explicitly. The deposition rate of Mg controlled by a quartz micro-balance was $0.5$\,ML/min. After MgO:N deposition, the samples are annealed at 1100\,K for 10\,min. Figure \ref{fig_0}(a) shows a scanning tunneling microscopy (STM) image of 7\,ML MgO$_{0.973}$N$_{0.027}$. It exhibits islands on top of an MgO:N wetting layer. From comparison of the coverage determined by the quartz balance and the volume of the MgO:N islands, we estimate the wetting layer thickness to 1-2 ML.
We checked crystalline quality and chemical purity of Mo(100) and the MgO:N films by low energy electron diffraction (LEED) and Auger electron spectroscopy (AES).
The nitrogen concentration $x$ within the MgO films was determined by AES at primary electron energy of 1\,keV being sensitive to the upper 2\,nm (9 to 10 ML) only \cite{Akkerman}. The atomic concentration $x_i$ of element $i$ within
homogeneous films is calculated from the AES peak height $Y_i$ shown in Fig. \ref{fig_1}(a) according to:\cite{Auger}

\begin{equation}
  x_i = \frac{Y_i/S_i}{\sum\limits_{a}^{} Y_a/S_a},
\end{equation}
where $S_i$ denotes the normalized sensitivity factor and $a$ sums over all relevant elements.\\
%
%N concentration
\begin{figure}
  \includegraphics[width=7.7cm]{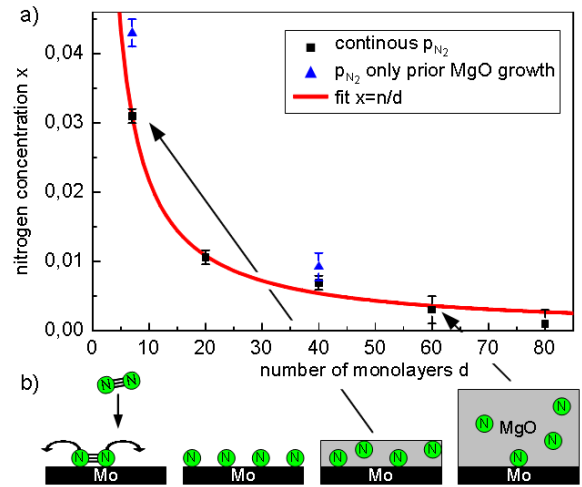}
  \caption{\label{fig_2}(Color online) a) Nitrogen concentration $x$ as a function of monolayers $d$; different symbols mark different preparation methods; fit curve assumes a constant amount of nitrogen atoms (see text). (b) Sketch of N$_2$ incorporation: left: N$_2$ dissociation on Mo(001); right: incorporation of N into MgO leaving the amount of N independent of MgO thickness.}
\end{figure}
The N concentration $x$ depends on deposition temperature $T_{\rm D}$ as shown in Fig.\ref{fig_1}(b) for a 7 ML MgO:N film. Up to $x=6$\,\% is achieved at $T_{\rm D}=850$\,K indicating a more effective dissociation of N$_2$ at higher $T_{\rm D}$. Notice that $x=6$ \% at 7 ML is still far below the amount of N expected from a full N coverage of Mo(001).
Next, we prepared MgO:N films with a thickness of up to 100\,ML at $T_{\rm D}=900$ K. The nitrogen amount within the MgO:N decreased dramatically with film thickness as shown in Fig. \ref{fig_2}. A 60\,ML thick MgO:N film contained only 0.3\,\% nitrogen in comparison to 3.2\,\% at a thickness of 7\,ML. Assuming homogeneous distribution of N, this implies that the amount of N in both films is roughly the same evidencing that N$_2$ is dissociated on the Mo(100) surface only. We assume that during MgO growth or annealing, the atomic N from the surface is incorporated into the MgO film. To support this scenario, we fit the data assuming a constant areal concentration $n$ of of N atoms:
\begin{equation}
  x=\frac{n}{d},
  \label{eq1}
\end{equation}
where $d$ denotes the number of MgO ML and $n$ represents the concentration of N atoms with respect to the sum of N and O atoms. If we assume that all N atoms are  substitutionally incorporated into the first monolayer of MgO:N, the fit curve plotted in figure \ref{fig_2}(a) shows excellent agreement with the measured data points using $n=21.6\,\%$. Assuming interstitial impurities, i.e. dumbbells of NO\cite{Pesci}, equation (\ref{eq1}) has to be slightly modified and results in $n=18.2\,\%$.
Nearly the same $n$ can be achieved by another preparation method: Prior to the deposition of Mg in a pure O$_2$ environment, we exposed the Mo crystal to N$_2$ ($p=5\cdot 10^{-6}$\,mbar) at 300 K for $10$ min. Afterwards, we grow MgO without N$_2$ at $T = 900\,$K leading to very similar N concentrations as shown in Fig. \ref{fig_2}(a). Thus, obviously, the dissociation of
N$_2$ takes place at the Mo(001) only.
Finally, a third preparation has been performed: 10\,ML of pristine MgO were grown first at $p_{\rm N_2}< 10^{-11}$ mbar. Subsequently, 10 ML MgO:N are deposited at $T_{\rm D}=300$ K and $p_{\rm N_2}=5\cdot 10^{-6}$ mbar. No nitrogen was found in the sample, i.e. $x<0.2$ \%. Thus, if Mo is covered by MgO, the catalytic effect of the substrate is inhibited.\\
\begin{figure}
  \includegraphics[width=8cm]{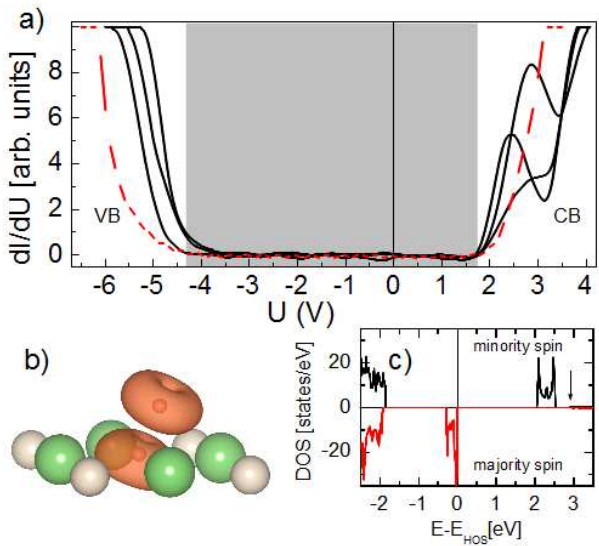}
  \caption{\label{fig_3}(Color online) a) $dI/dU(U)$ spectra measured by STS on a 11 ML high MgO$_{0.96}$N$_{0.04}$ island ($U_{\text{stab}} = 3$\,V, $I_{\text{stab}} = 0.5$\,nA, $U_{\text{mod}} = 40$\,mV) at several positions (straight lines) and on a 11 ML thick pristine MgO island (dashed line); average film thickness in both cases: 7 ML.
 b) Calculated charge density of the unoccupied N-induced states of a N-N dimer at the surface. Large green and small white spheres show Mg and O respectively.
 c) Calculated density of states (DOS) for N-N dimer at MgO surface, HOS: highest occupied state; all states between  -1.2 eV and 3 eV are N-induced; CBM is marked by arrow.}
\end{figure}
Figure \ref{fig_3} compares STS curves of 11 ML films of MgO$_{0.96}$N$_{0.04}$ and undoped MgO.\cite{Grob} Doping by N leads to a shift of the Fermi level $E_{\rm F}$ towards the valence band by about 1 eV and an additional peak with maximum at $0.3$ eV below the conduction band minimum (CBM). This is observed for three different N concentrations, i.e. MgO thicknesses, as probed by STS. The peak energy varies laterally by $\pm 0.3$ eV. Density functional (DFT) calculations of bulk MgO with substitutional N predict occupied p-levels close to the valence band maximum (VBM) and an unoccupied, spin polarized level within the middle of the band gap, if self interaction correction is included.\cite{Droghetti,Slipukhina}\\
This disagrees with our experiment.
Since surfaces are not included in these calculations, we performed first-principles DFT calculations including the surface
within the spin-polarized generalized gradient
approximation~\cite{PBE1996} using projector
augmented-wave potentials as implemented in the Vienna Ab initio Simulation Package (VASP).~\cite{Kresse}
Correlation effects on
the p-shells of N dopants are accounted for by the DFT+U scheme
in Dudarev's approach~\cite{Dudarev} with an on-site effective Coulomb parameter
U$_{\rm eff}$=3.4 eV. A kinetic energy cutoff of 500 eV and a $6\times6\times1\ \Gamma$-centered $k$-point mesh
was used. The supercell consists of 9 atomic layers MgO(001) using the experimental lattice parameter and  a 16~\AA\ thick layer
of vacuum. All atomic positions as well as the thickness of the MgO slab were fully relaxed.\\
Three different configurations with N atoms in the surface layer have been calculated: ($i$) one N-atom substituting an oxygen, ($ii$) one N-atom at the interstitial site and ($iii$) an N-N dimer with one N atom being substitutional and the other at the nearest interstitial site. In case ($ii$), the calculation was initiated with an N interstitial, but the system relaxed to a configuration where N and O exchanged their places, i.e. N ended up substitutionally and O interstitially.
However, the N derived p-states for case ($i$) and ($ii$) are found within 2.2 eV above VBM very similar as for MgO bulk.\cite{Droghetti,Slipukhina} We conclude that a more complex N structure as, e.g., the N-N dimer at the surface shown in Fig. \ref{fig_3}(b) and exhibiting unoccupied N-type p-states close to CBM (Fig. \ref{fig_3}(c)), is responsible for the observed peak.\\
In conclusion, we prepared thin films of N-doped MgO with an N-concentration up to $6\,\%$ by using the catalytic effect of Mo(001) for N$_2$ dissociation.
Compared with pristine MgO, an additional state close to the conduction band minimum has been observed by STS which could not be attributed to simple N impurity
configurations.\\
Helpful discussions with P. Mavropoulos and S. Parkin as well as financial support by SFB 917-A3 and HGF\_YIG VH-NG-409 are gratefully acknowledged.


\begin{thebibliography}{5}
\bibitem{Yang}
C. H. Yang, Ph. D. thesis, Stanford university, Stanford 2010.
\bibitem{Ming} L. Chun-Ming {\it et al.},
Chin. Phys. B {\bf 20}, 047505 (2011).
\bibitem{HfO2}
M. Ventakesan, C. B. Fitzgerald, and J. M. D. Coey, Nature {\bf 430}, 630 (2004);
J. M. D. Coey {\it et al.}, Phys. Rev. B {\bf 72}, 024450 (2005); M. Khalid {\it et al.}, Phys. Rev. B \textbf{80}, 035331 (2009); N. H. Hong {\it et al.}, Phys. Rev. B {\bf 73}, 132404 (2006), D. Gao {\it et al.},
 J. Phys. Chem C {\bf 114}, 11703 (2010).
\bibitem{MgO}
 C. Moyses Araujo {\it et al.}, Appl. Phys. Lett. {\bf 96}, 232505 (2010); C. Martinez-Boubeta {\it et al.}, Phys. Rev. B
 {\bf 82}, 024405 (2010).
 \bibitem{Esqui}
 P. Esquinazi {\it et al.}, Phys. Rev. Lett. {\bf 91}, 227201 (2003); S. Talapatra {\it et al.}, Phys. Rev. Lett. {\bf 95}, 097201 (2005).
\bibitem{Pan}
 H. Pan {\it et al.}, Phys. Rev. Lett. {\bf 99}, 127201 (2007).
 \bibitem{Yu}
 C. F. Yu {\it et al.}, J. Phys. D: Appl. Phys {\bf 40}, 6497 (2007); S. Chawla {\it et al.}, Phys. Rev. B {\bf 79}, 125204 (2009);
 J. Appl. Phys. {\bf 106}, 113923 (2009)
 Y. Ma {\it et al.}, IEEE Trans. Magn. {\bf 46}, 1338 (2010).
 \bibitem{Li}
 X. L. Li {\it et al.}, IEEE Trans. Magn. {\bf 46}, 1382 (2010); X. G. Xu {\it et al.}, Appl. Phys. Lett. {\bf 97}, 232502 (2010);
 J. B. Yi {\it et al.}, Phys. Rev. Lett. {\bf 104}, 137201 (2010); D. Gao {\it et al.},  J. Phys. Chem {\bf 14}, 13477 (2010);
 V. Bhosle and J. Narayan, Appl. Phys. Lett. {\bf 93}, 02192 (2008).
 \bibitem{Kenmochi}
L.S. Efimov, S.Yunoki, and G. A. Sawatzky, Phys. Rev. Lett. {\bf 89}, 216403 (2002);
K. Kenmochi {\it et al.}, Jpn. J. Appl. Phys {\bf 43},
L 934 (2004); K. Kenmochi {\it et al.}, J. Phys. Soc. Jp.
{\bf 73}, 2952 (2004); V. A. Dinh {\it et al.}, J. Phys. Soc. Jpn. {\bf 75}, 093705 (2006);
I. S. Efimov {\it et al.}, Phys. Rev. Lett. {\bf 98}, 137202 (2007).
\bibitem{Mavropoulos} P. Mavropoulos, M. Le\v zai\' c, and S. Bl\"ugel, Phys. Rev. B \textbf{80}, 184403 (2009).
\bibitem{Droghetti} A. Droghetti, C.D. Pemmaraju, and S. Sanvito, Phys. Rev. B \textbf{78}, 140404 (2008); V. Pardo and W. E. Picket, Phys. Rev. Lett. {\bf 78}, 134427 (2008); H. Wu, A. Stroppa, S. Sakong, S. Picozzi, M. Scheffler, P. Kratzer,
Phys. Rev. Lett. \textbf{105}, 267203 (2010). 
\bibitem{Slipukhina} I. Slipukhina {\it et al.}, Phys. Rev. Lett. {\bf 107}, 137203 (2011).
\bibitem{Gu} B. Gu {\it et al.}, Phys. Rev. B {\bf 79}, 024407 (2009).
\bibitem{Waser}R. Waser {\it et al.}, Adv. Mat. {\bf 21}, 2632 (2009).
\bibitem{Pesci} M. Pesci {\it et al.}, J. Phys. Chem. C \textbf{1141}, 1350 (2010).
\bibitem{Gallagher} M.C. Gallagher {\it et al.}, Thin Solid Films \textbf{445}, 90 (2003).
\bibitem{Benedetti} S. Benedetti {\it et al.}, Chem. Phys. Lett. \textbf{430}, 330 (2006).
\bibitem{Grob}
C. Pauly {\it et al.}, Phys. Rev. B {\bf 81},125446 (2010).
\bibitem{Chatt} J. Chatt {\it et al.}, Nature \textbf{224}, 1201 (1969).
\bibitem{Alducin} M. Alducin {\it et al.}, Phys. Rev. Lett. \textbf{97}, 056102 (2006).
\bibitem{Clavenna} L.R. Clavenna and L.D. Schmidt, Surf. Sci. \textbf{22}, 365 (1970).
\bibitem{Rettner} C.T. Rettner, E.K. Schweizer, and H. Stein, J. Chem. Phys. \textbf{93}, 1442 (1990).
\bibitem{Bode} M. Bode {\it et al.}, Surf. Sci. \textbf{601}, 3308 (2007).
\bibitem{Akkerman} A. Akkerman {\it et al.}, phys. stat. sol. (b) \textbf{198}, 769 (1996).
\bibitem{Auger} S. Mroczkowski and D. Lichtmann, J. Vac. Sci. Technol. A \textbf{3}, 1860 (1985).
\bibitem{PBE1996}
J. P. Perdew {\it et al.}, Phys. Rev. Lett. {\bf 77}, 3865 (1996).
\bibitem{Kresse}
G. Kresse {\it et al.}, Phys. Rev. B {\bf 47}, 558 (1993); {\bf 54}, 11169 (1996); {\bf 59}, 1758 (1999); P. E. Bl\"ochel,
Phys. Rev. B {\bf 50}, 17953 (1994).
\bibitem{Dudarev}
S. L. Dudarev {\it et al.}, Phys. Rev. B {\bf 57}, 1505 (1998).
\end{thebibliography}
\end{document}